\documentclass[fleqn,usenatbib]{mnras}

\usepackage{newtxtext,newtxmath}

\usepackage[T1]{fontenc}

\DeclareRobustCommand{\VAN}[3]{#2}
\let\VANthebibliography\thebibliography
\def\thebibliography{\DeclareRobustCommand{\VAN}[3]{##3}\VANthebibliography}

\usepackage{graphicx}	

\newcommand\egdrthree{\textit{Gaia}~EDR3 }
\newcommand\gdrtwo{\textit{Gaia}~DR2 }

\def\ltsima{$\; \buildrel < \over \sim \;$}
\def\simlt{\lower.5ex\hbox{\ltsima}}
\def\gtsima{$\; \buildrel > \over \sim \;$}
\def\simgt{\lower.5ex\hbox{\gtsima}}
\def\kms{{\rm\,km\,s^{-1}}}

\def\masyr{{\rm\,mas/yr}}
\def\muasyr{{\rm\,\mu as/yr}}
\def\kpc{{\rm\,kpc}}


\def\deg{^\circ}

\def\s{\ifmmode \widetilde \else \~\fi}
\def\={\overline}

\def\spose#1{\hbox to 0pt{#1\hss}}

\def\lta{\mathrel{\spose{\lower 3pt\hbox{$\mathchar"218$}}
     \raise 2.0pt\hbox{$\mathchar"13C$}}}
\def\gta{\mathrel{\spose{\lower 3pt\hbox{$\mathchar"218$}}
     \raise 2.0pt\hbox{$\mathchar"13E$}}}
\def\Dt{\spose{\raise 1.5ex\hbox{\hskip3pt$\mathchar"201$}}}    
\def\dt{\spose{\raise 1.0ex\hbox{\hskip2pt$\mathchar"201$}}}    

\def\dotsfill{\leaders\hbox to 1em{\hss.\hss}\hfill}

 \title[M31 proper motion with \egdrthree]{The proper motion of Andromeda from \egdrthree:\\ confirming a nearly radial orbit}

\author[J.-B. Salomon et al.]{
J.-B. Salomon,$^{1,2,3}$\thanks{E-mail: jean-baptiste.salomon@utinam.cnrs.fr}
R. Ibata,$^{4}$
C. Reylé,$^{3}$
B. Famaey,$^{4}$
N. I. Libeskind$^{2,5}$
A.W. McConnachie$^{6}$
\newauthor {and Y. Hoffman,$^{1}$}
\\
$^{1}$Racah Institute of Physics, Hebrew University, Jerusalem 91904, Israel\\
$^{2}$Leibniz-Institut für Astrophysik Potsdam, An der Sternwarte 16, D-14482 Potsdam, Germany\\
$^{3}$Institut UTINAM, CNRS UMR6213, Universit\'e Bourgogne Franche-Comt\'e, OSU THETA Franche-Comt\'e-Bourgogne,\\
   Observatoire de Besançon, 41 bis avenue de l'Observatoire, BP 1615, 25010 Besan\c{c}on C\'edex, France\\
$^{4}$Observatoire astronomique de Strasbourg, Université de Strasbourg, CNRS, 11 rue de l’Université, 67000 Strasbourg, France\\
$^{5}$University of Lyon, UCB Lyon 1/CNRS/IN2P3, IPN Lyon, F-69622 Villeurbanne, France\\
$^{6}$NRC Herzberg Astronomy and Astrophysics, 5071 West Saanich Road, Victoria, BC V9E 2E7, Canada
}

\date{Accepted 2021 August 2. Received 2021 July 31; in original form 2020 December 18}

\pubyear{2020}

\begin{document}
\label{firstpage}
\pagerange{\pageref{firstpage}--\pageref{lastpage}}
\maketitle

\begin{abstract}
We present an analysis of the proper motion of the Andromeda galaxy (M31), based on the Early Third Data Release of the {\it Gaia} mission. We use the {\it Gaia} photometry to select young blue main sequence stars, and apply several quality cuts to obtain clean samples of these tracers. After correcting the proper motion measurements for the internal rotation of the M31 disk motion, we derive an apparent motion of $52.5 \pm 5.8\muasyr$ with respect to the {\it Gaia} reference frame, or $61.9 \pm 9.7\muasyr$ after applying a zero-point correction determined from quasars within $20\deg$ from M31 and a correction from systemic biases. Accounting for the Solar reflex motion we deduce a relative velocity between Andromeda and the Milky way (in a non-rotating frame at the current location of the Sun) of $42.2 \pm  39.3 \kms$ along right ascension ($40.0 \pm  39.3 \kms$ along galactic longitude) and $-59.4 \pm  30.3 \kms$ along declination ($-60.9  \pm  30.3 \kms$ along galactic latitude), with a total transverse velocity of $V_{\rm trans} = 82.4 \pm  31.2 \kms$. These values are consistent with (but more accurate than) earlier Hubble Space Telescope measurements that predict a future merger between the two galaxies. We also note a surprisingly large difference in the derived proper motion between the blue stars in M31 and samples of red stars that appear to lie in that galaxy. We propose several hypotheses to explain the discrepancy but found no clear evidence with the current data to privilege any one of them.
\end{abstract}

\begin{keywords}
galaxies: kinematics and dynamics; Local Group -- proper motions
\end{keywords}



\section{Introduction}\label{sect:intro}

The Andromeda galaxy (M31) and the Milky Way (MW) are the dominant gravitational components of our nearby environment in the local Universe. These two giant spiral galaxies and their cohort of satellites, among which the Triangulum galaxy (M33) and the Magellanic Clouds (MC) are the most massive ones, form what is called the Local Group (LG). Often considered as the close cousin of the MW, due to their resemblance in terms of mass, shape and evolutionary stage, M31 is located at a distance of $785\pm25\kpc$ \citep{McConnachie05}.

This structural and spatial proximity makes Andromeda the preferred observational target for understanding the formation and evolution of our own Galaxy in its local environment. However, it has also become clear that the similarities between the two spirals are actually limited when their properties are studied in more detail. For example, while the Milky Way experienced its last major merger about ten billion years ago \citep{Helmi18}, the recent history of M31 seems to be more disturbed, as testified by the presence of the Giant Stellar Stream \citep{Ibata01,McConnachie03} or its more disturbed stellar disk \citep{Hammer18}.

The numerous discoveries of stellar streams around the two spiral galaxies \citep[e.g.,][]{Chapman08, Martin14}, remind us that the build-up of a galaxy is intimately linked to the environment in which it takes place. Thus, in order to understand the history of the formation of the Milky Way and M31, it is crucial to place the galaxies in their environmental context, meaning the LG.

However, there are many peculiarities in the LG that can perhaps call into question our understanding of galaxy formation. One example is the lopsided distribution of dwarf galaxies in the Local Group, where satellites tend to be located between the two giant spirals \citep{Conn12, Libeskind16}. This distribution is even less uniform when observing the phase-space distribution of satellites around the Milky Way \citep{Pawlowski12} and M31 \citep{Ibata13}, which form highly flattened and apparently rotating structures. A majority of the satellite galaxies contained within these planes indeed appear to have the same co-rotational kinematics \citep{Pawlowski20}.

We do not know whether these peculiarities are intrinsic, specific to the LG, or whether our interpretation and modelling of galaxy formation processes are inaccurate. While the existence of galaxy pairs is not unique in the Universe, it is sufficiently rare in our close proximity to prevent us from statistically studying the detailed physical properties of such configurations, due to the lack of observational data. The fundamental question is therefore whether our place of residence is a peculiar spot, which by coincidence is a statistical exception, or whether our understanding of galaxy formation is still lacking some fundamental ingredients.

To try to shed some light on these issues, it is necessary to characterise the properties of the LG as accurately as possible. One of the most fundamental aspects to uncover the dynamical evolution of the LG is the relative velocity between its two main galaxies. This velocity vector can tell us a lot about the role played by the group having only two major galaxies. Does having such a pair impact the evolution of either galaxy in the pair, or did they evolve independently of each other? If some coupling exists, is the group gravitationally linked or bound?

This motion is, first of all, decisive in the calculation of the mass of the LG as well as the position and velocity of its barycentre via the so-called timing argument \citep{Kahn59, LyndenBell81, Penarrubia14}. These quantities allow us to better place the LG in its cosmological context. At the boundary of the LG, the gravity it generates competes with cosmological expansion. The surface (most of the time modelled as a sphere) where the two forces exactly compensate each other is the zero velocity surface \citep{Tully15}. With a good knowledge of the parameters of the barycentre, it is then possible to place the LG in the cosmic web and thus have a refined view of the way in which accretions occur \citep{Courtois13}. 

A good knowledge of the relative velocity between M31 and the MW allows to refine our knowledge about their orbits, and thus to determine more precisely the gravitational field of the LG in its vicinity but also within it \citep{Peebles17}. It may then become possible to decide between several hypotheses of formation such as a first encounter between M31 and the MW, or a second passage after a previous interaction, or a past important accretion in one of the galaxies that could have affected the environment of the other \citep{Hammer13}. Knowledge of the orbits even makes it possible to test different gravitational models \citep{Zhao13, Banik16, Carlesi17, Bilek18}, via for example the timing argument \citep{Benisty19}.
Moreover, knowing precisely the relative velocity of the two giants, allows at the same time to model precisely the orbits of their satellites, especially when their own motion is known as in the case of M33 \citep{Patel17, Semczuk18,TepperGarcia20}. 

And of course, the relative velocity of the two spirals tells us directly about their future, will there be a merger or not? Armed with the velocity measurement, detailed modelling is then possible to determine what the properties of the different components will become \citep{Hoffman07, Cox08, vdM12b, Schiavi20}.

Nonetheless, while the radial velocity (along the line of sight) of the Andromeda galaxy has been precisely known for nearly a century \citep{Slipher1913}, it is only within the last decade that we have been able to begin to tackle its proper motion.
The first direct measurement of the transverse velocity of M31 was made by \cite{Sohn12} using data from the Hubble Space Telescope (HST). The latter observed three fields in and around Andromeda over a period of 5 to 7 years baseline. As the HST fields of view are small, it is necessary to model the internal dynamics of M31 to derive a constraint on the global motion of the galaxy. The result is therefore highly dependent on the model \citep{vdM12a}. The weighted final values in the heliocentric frame are $163\kms$ in the east direction in RA and $-117\kms$ towards the north in declination, with uncertainties of $\sim 45\kms$. This implied a rather surprising nearly radial orbit between the two spirals.

The second direct measurement was carried out by \cite{vdM19} with the second {\it Gaia} data release (\gdrtwo) catalogue. Although this catalogue has large random uncertainties, of the order of $1000\kms$ for the targeted sources, having the entire field of the galaxy with about 1000 stars fairly uniformly distributed makes the proper motions calculation possible, $\sim 240 \kms$ towards the east and $\sim -210\kms$ towards the north in the heliocentric referential. The uncertainties being $\sim 120\kms$.

The transverse velocity measurement of M31 has also been tentatively derived indirectly, by analysing the perspective motion. Under the assumption that the satellite system follows the same global dynamics as the central galaxy, it is possible to calculate, statistically, the overall velocity of the Andromeda system. The method is based on fitting the radial velocities of the satellites, each being on a different line of sight, and hence with different projections of the bulk motion. With a sample of 24 satellites (of which 5 are considered as LG satellites, and 2 have an already known proper motion), \cite{vdM08} (vdM08) estimated a velocity of $97\kms$ towards the east and $45 \kms$ towards the north (revised in \cite{vdM12a}) with uncertainties of about $38\kms$. \cite{Salomon16} (S16) measured M31's proper motion using 39 satellites with distance constraints, obtaining values of $-111\kms$ and $99\kms$ with uncertainties of $\sim 65\kms$. If only those galaxies outside of the co-rotating plane of satellites are considered (26), the proper motion was found to be $-78\kms$ and $1\kms$ with uncertainties of $\sim 68\kms$.

The diversity of values makes the relative motion of the two spirals quite equivocal. In the face of the amplitude of the uncertainties, all these values are nevertheless in agreement with each other at two sigma. It should be noted that while the values of HST, vdM08 and S16 (without the plane) are compatible with a radial trajectory from M31 towards the Milky Way, the values of \gdrtwo and S16 (39 satellites) deviate significantly from a future head-on collision. The latter value even implies that the pair of spirals would not be gravitationally bound, which would drastically change the nature of the LG. The heterogeneity of the values indicates above all that each of the methods is dominated by systematic biases. 
This is why we have undertaken to revisit the direct determination of the transverse velocity of M31 by taking advantage of the exquisite new third {\it Gaia} early data release (\egdrthree) Catalogue \citep{Brown21}.

The article is organised as follows. In section~\ref{sect:selection}, we present the selection of our sample, with the different cuts used in order to best study the sources of M31. Then, in section~\ref{sect:method}, the modelling of the disk is described and the method is validated. The results are presented in section~\ref{sect:results}. Finally, in section~\ref{sect:ccl} we discuss the implications that might arise from this new value and draw the conclusions.

\section{Selection of sources from \egdrthree} \label{sect:selection}

\subsection{Generic cleaning up}\label{sect:generic}

The \egdrthree catalogue \citep{Brown21}, based on a continuous observation time of 34 months, provides precise photometry and astrometry for about 1.8 billion sources. Uncertainties on proper motions are divided by a factor of about two with respect to the \gdrtwo catalogue \citep{Brown21}. Several reference articles provide details on the validation of the catalogue \citep{Fabricius21}, on the photometry \citep{Riello21} and on the astrometry \citep{Lindegren21}. In the following, we make use of the \egdrthree data, by applying various parameter cuts, to target sources which are likely to be bona fide bright stars in M31. The selection process with the corresponding number of sources retained is summarised in Table~\ref{tab:cut_sum}.

\begin{figure}
	\includegraphics[width=\columnwidth]{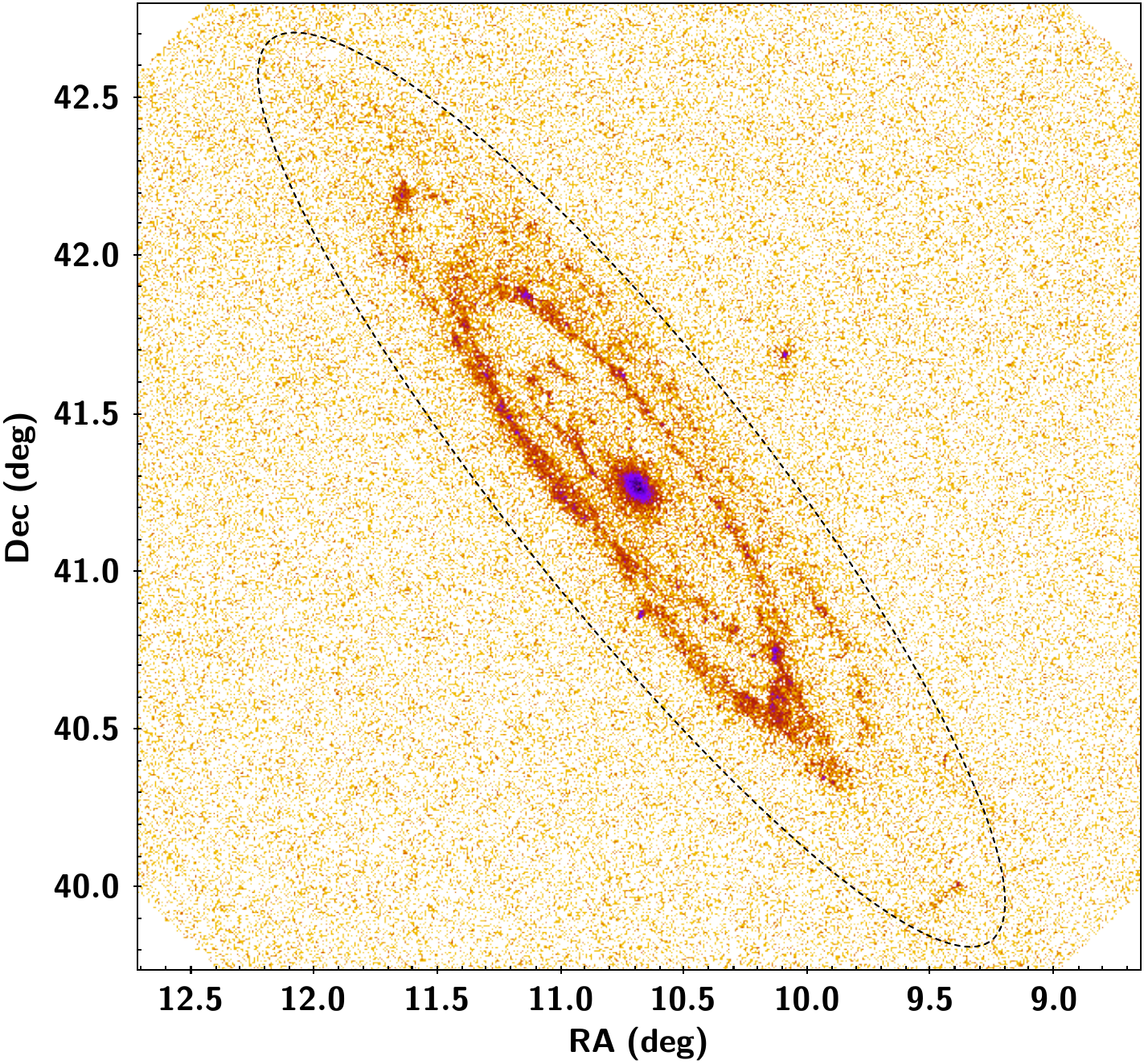}
    \caption{Selection from the \egdrthree catalogue of all sources within a radius of 2 degrees from the centre of Andromeda (204 386 sources). The thin black dashed line represents our central spatial selection described at the end of section~\ref{sect:generic}.}
    \label{fig:m31_full_edr3}
\end{figure}

We first query the {\it Gaia} archive\footnote{https://gea.esac.esa.int/archive/} to select all sources within a radius of two degrees centred on M31's position (RA, Dec) = ($10.68\deg$, $41.27\deg$) \citep{Evans10}. In this area, the \egdrthree catalogue contains 204 386 sources which are depicted in Figure~\ref{fig:m31_full_edr3}.
The galaxy is already visible as clear over dense regions on the sky. The bulge, the satellite galaxy NGC 205 and the ring of active star formation are especially noticeable. A second larger ring-like shape can also be distinguished as well as some characteristic arm features. 

Given the fact \egdrthree does not provide accurate parallaxes at the distance of Andromeda (many of the parallaxes are actually negative), we remove all of the foreground sources within a distance of 10 kpc from the Sun, with one sigma confidence interval with the selection $\varpi - \sigma_{\varpi} < 0.1$ where $\varpi$ is the parallax in milli-arcsec. This selection allows us to eliminate certain foreground sources, while being tolerant on sources whose error on the parallax would permit larger distances. It also allows us to remove sources with no parallax measurement. We thereby retain 73 992 sources.

At this early stage, we also take conservative restrictions in the colour magnitude diagram (CMD) with $G > 16$ and $-1 < G_{BP} - G_{RP} < 4$. This ensures at the same time that the remaining 72 413 sources have photometric information.

Even if greatly improved compared to {\it Gaia} DR2, the photometric data of \egdrthree still suffers from excess flux (ratio of $G_{\rm BP}$ and $G_{\rm RP}$ bands to $G$), especially in very dense regions and towards red colours. We follow the recommendations made by \cite{Riello21} to calculate the corrected excess factor (C$_{\rm corr}$) (see their Equations 6, 18 and Table 2). We select sources having a scatter smaller than 3 $\sigma$ with:
\begin{equation}
    |C_{\rm corr}|> 3 \big( 5.9898.10^{-3} \times 8.817481.10^{-12} \times G^{7.618399} \big)
\end{equation}
After this procedure 56 032 sources remain in the sample.

In order to conserve good astrometric solutions for the remaining sources, we make use of the renormalised unit weight error factor provided in the \egdrthree archive as {\texttt{ruwe}} which is particularly useful for unresolved sources. Following the prescription of \cite{Lindegren21}, we inspect the data for strong deviations in the distribution of this factor on our sample. The distribution is approximately normal, almost centred on 1 and presents a tail towards high values. Certain sources have a large excess, beyond the visible as a break in the distribution above $1.3$. Consequently, sources with a {\texttt{ruwe}} > 1.3 are rejected from our sample.
We finally clean up the sample from potential binary sources or at least detected as such by the {\it Gaia} reduction algorithm in applying the following cuts: {\texttt{ipd\_gof\_harmonic\_amplitude}} < 0.1 and {\texttt{ipd\_frac\_multi\_peak}} < 2 \citep{Fabricius21}.
This leaves us with 53 768 sources.

We finally apply a geometric spatial selection to better follow the shape of the projected M31 disk on the sky (the thin black dashed ellipse on Figure~\ref{fig:m31_full_edr3}). For this crude selection, the angular size of the disk adopted is $1.8\deg$, appearing as an ellipse in the sky viewed form perspective angles of $i = 77.5\deg$ for the inclination and ${\rm PA} = 37.5\deg$ for the position angle, an average value from the literature (see e.g. \citet{Chemin09, Corbelli10} reducing the number of sources to 18 173.

In order to optimally select M31 stars in the CMD and to better evaluate the amount of contaminants, we build two other samples, also based on geometrical cuts. The elliptical selection on the sky is conserved but shifted, for one sample slightly towards the north-west (``out~1''), for the other one slightly towards the south-east (``out~2'') in such a way so that the three ellipses do not overlap but are still contained in the initial sample (see top panel of Figure~\ref{fig:m31_ellipse_CMD_edr3}). The north-west comparison field contains 7798 sources and the south-east, 7083.

\begin{table}
	\centering
	\caption{Summary of the various cuts applied to the \egdrthree catalogue to build the M31 samples with the corresponding number of sources retained. The last three rows indicate the final selections with the number of sources within the central ellipse (``in''), the north-west ellipse (``out~1'') and the south-east ellipse (``out~2'').}
	\label{tab:cut_sum}
	\begin{tabular}{lc} 
		\hline
		Cuts & Remaining sources\\
		\hline
		2$\deg$ from M31 centre         & $204 386$ \\
		Distance > 10 kpc               & $73 992$  \\
		CMD conservative restrictions   & $72 413$  \\
		Flux excess correction          & $56 032$  \\
		Resolved and non-binary sources & $53 768$  \\
		\hline
		Spatial selection (in)          & $18 173$  \\
		Spatial selection (out 1)       & $7 798$   \\
		Spatial selection (out 2)       & $7 083$   \\
		\hline
	\end{tabular}
\end{table}

\subsection{Sample selections}\label{sect:subsample}
Following the initial parameter cuts described above, hereafter we present several subsequent selections, based on various hypotheses. When comparing final results derived with these samples, this will enable us to better understand biases that stem from each of the selections and to better assess the derived proper motions.

\subsubsection{Fiducial sample (Blue)}
In this section we present in detail the way our favoured sample is constructed. Other samples will be quickly introduced in the following subsections. This sample, which we also refer to as the fiducial sample, contains blue bright stars.

\begin{figure}
	\includegraphics[width=\columnwidth]{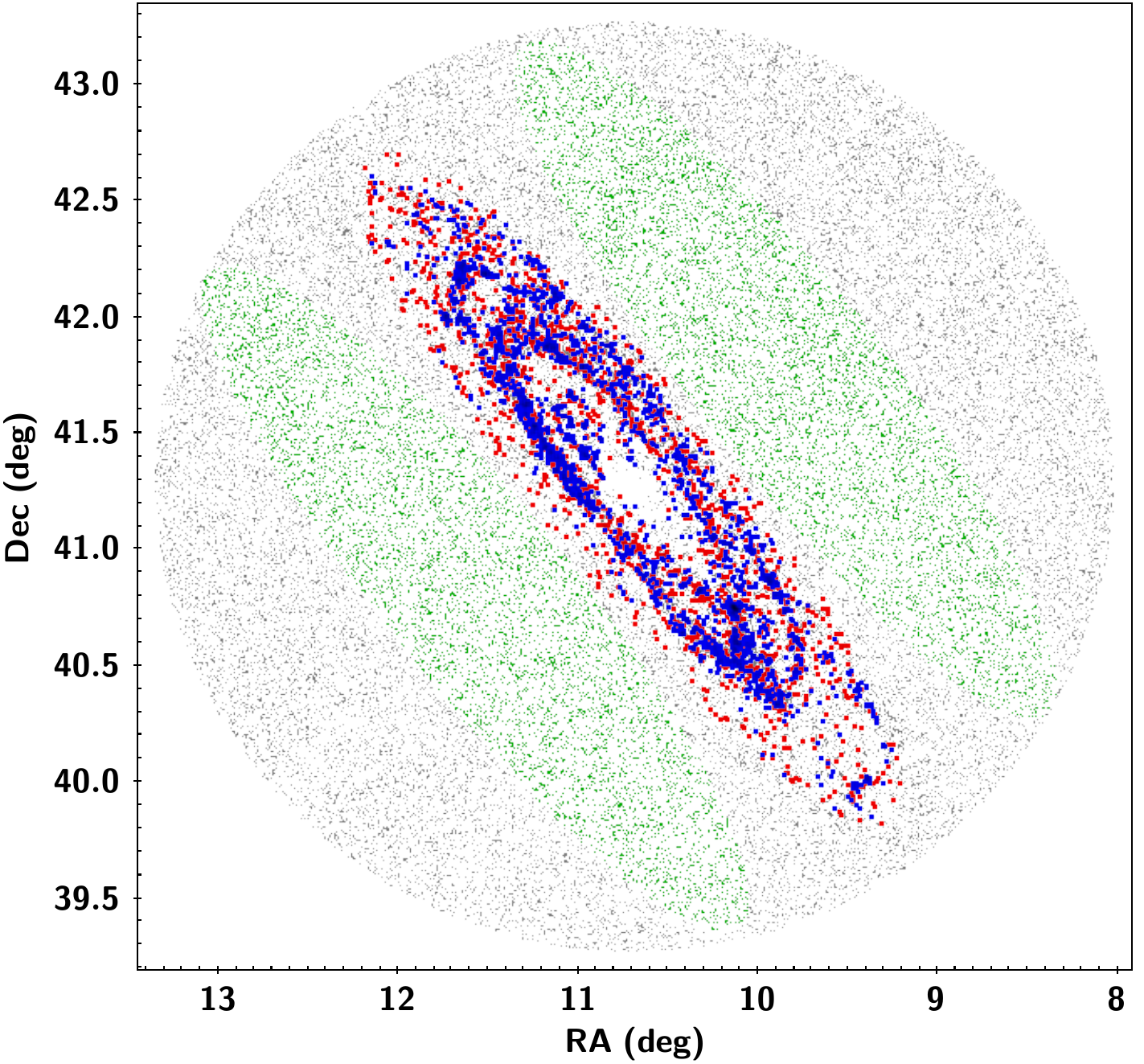}
	\includegraphics[width=\columnwidth]{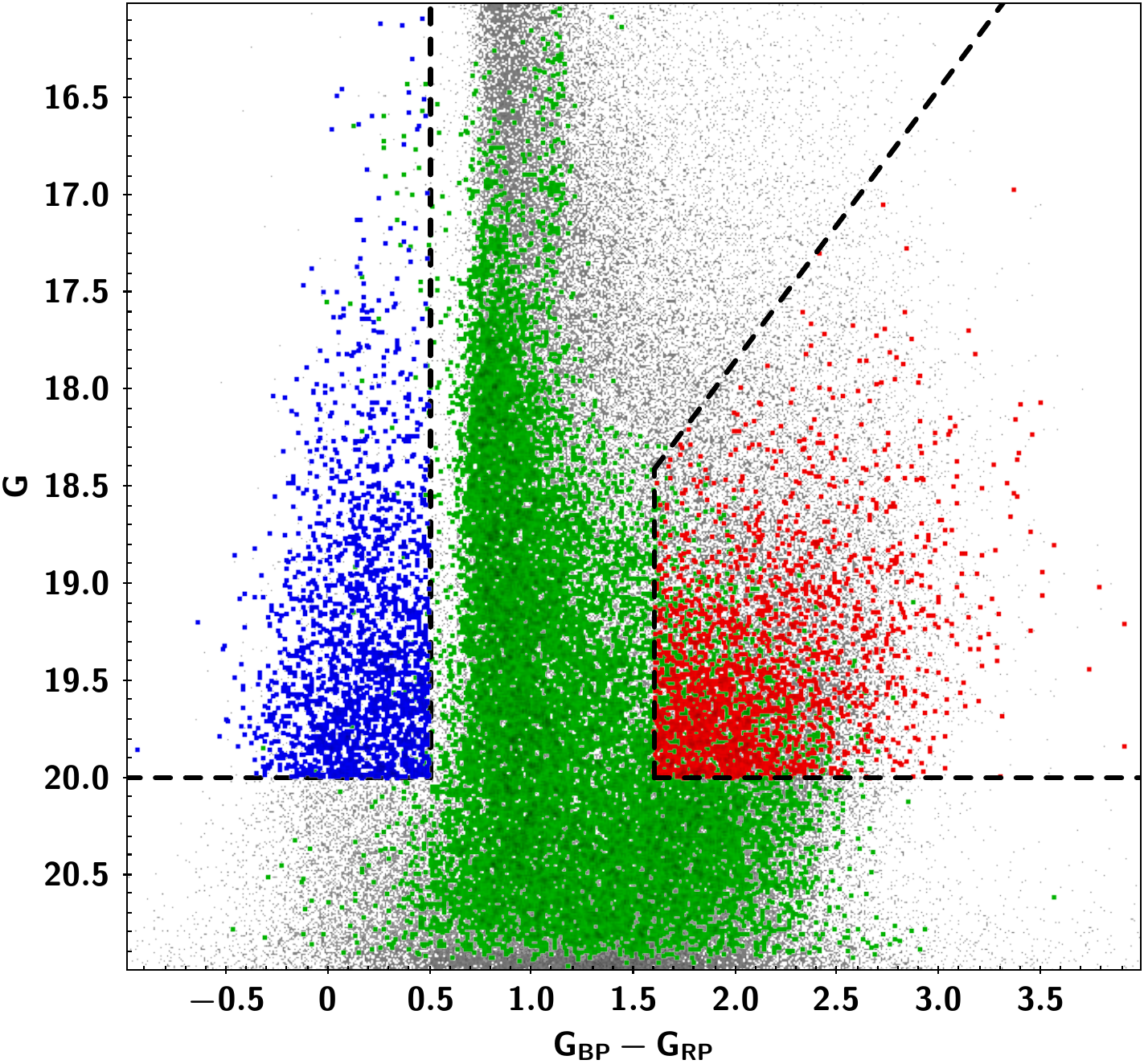}
    \caption{Top panel: sky view of the different selections. In grey dots: full sample from \egdrthree catalogue of all sources within a radius of 2 degrees from the centre of Andromeda. Dots in blue, and red: samples from the \egdrthree catalogue after the main cuts and the ellipsoidal geometrical selection is applied. Dots in green: the comparison fields, shifted to the north-west and south-east. Bottom panel: CMD of the same sub-samples as in the top panel with an identical colour coding.}
    \label{fig:m31_ellipse_CMD_edr3}
\end{figure}

Despite the parameter cuts that have been applied, the ellipse centred on M31 still contains a certain amount of contaminants, as we can deduce from the large number of stars in the two adjacent comparison fields. To reduce the remaining parasitic sources, a selection is made in the CMD. It is clearly visible in the bottom panel of Figure~\ref{fig:m31_ellipse_CMD_edr3}, that 'field' sources are almost absent for $G_{\rm BP} - G_{\rm RP} < 0.5$. So we apply this conservative cut in colour to define a blue sample. Furthermore, given the dispersion in colour for faint stars, and once again to remain conservative, we apply a cut in magnitude, rejecting stars with $G > 20$. This `fiducial' selection leaves us with 1 919 sources for the central sample

Furthermore, as the two offset samples are juxtaposed to the field centred on M31, and the area covered in the sky is identical, it is straightforward to estimate the contamination in our fiducial M31 sample. The comparison fields harbour 38 sources for the south-east and 33 for the north-west, so the average amount of contamination is only about $1.8\%$. The ($1\sigma$) standard deviation in proper motion of these background stars is $(\rm{s}_{\alpha}, \rm{s}_{\delta}) = (1.640, 0.861) \masyr$.

\subsubsection{Red star sample (Red)}
\egdrthree also gives us the opportunity to investigate the proper motion of the very bright red sources in M31, which are clearly seen in the CMD (lower panel of Figure~\ref{fig:m31_ellipse_CMD_edr3}). These are probably red supergiants (and probably few asymptotic giant branch stars).
We select these sources by applying the CMD cuts $G_{BP}-G_{RP} > 1.6$ and $G > -1.4(G_{BP}-G_{RP}) + 20.65$, and limiting the faint sources to $G < 20$. After the geometric selection, the red sample contains 2083 sources with an estimated contamination of $33.15 $\% with a dispersion of $(\rm{s}_{\alpha}, \rm{s}_{\delta}) = (5.313, 4.423) \masyr$.

\subsubsection{Extended magnitude sample (B$_G$)}
Given the fact that \egdrthree extends to fainter magnitudes, we also take the opportunity to  relax the limit of $G < 20$, which brings us almost up to $G = 21$. This sample contains 4304 sources with a contamination level of 2.6\% and has $(\rm{s}_{\alpha}, \rm{s}_{\delta}) = (3.251, 1.711) \masyr$.

\subsubsection{Proper motion limited sample (B$_{\rm pm}$ and R$_{\rm pm}$)}

The values of the transverse velocity of M31 in the literature (see Section~\ref{sect:intro}) scan a wide range of values. They tell us, however, that it is excluded to at least one sigma to have a velocity greater than $400\kms$ in one direction in the heliocentric frame of reference. On the other hand, the rotation curve of the galaxy remains well below $300\kms$. A star belonging to Andromeda will thus have a velocity in each direction of less than $700 \kms$, equivalent to $\sim 190 \muasyr$. Stars that highly deviate from this limit are likely to be foreground stars or with extremely large proper motion uncertainties. To evaluate the impact of these sources on the derived transverse velocity of M31, we construct two new samples based on the `Blue' and `Red' selections above, limiting the proper motions to within three sigmas of the mean of the `fiducial' sample: $|\mu_{\alpha}-\mu_{\alpha}^{\rm fiducial}| < 0.19 + 3 \sigma_{\mu_{\alpha}}$ and $|\mu_{\delta}-\mu_{\delta}^{\rm fiducial}| < 0.19 + 3 \sigma_{\mu_{\delta}}$. The filtered blue sample contains 1861 sources with 0.9\% of contaminants and $(\rm{s}_{\alpha}, \rm{s}_{\delta}) = (0.260, 0.228) \masyr$. For the red sample, we thereby retain 1494 sources contaminated at a level of 2.7\% with $(\rm{s}_{\alpha}, \rm{s}_{\delta}) = (0.702, 0.624) \masyr$.

\begin{table}
	\centering
	\caption{Number of sources contained in each defined sample with their associated percentage level of estimated contamination.}
	\label{tab:sample_sum}
	\begin{tabular}{lcc} 
		\hline
		Sample &  Sources & Estimated contamination\\
		\hline
		Blue         & $1 919$       & $1.8 \%$  \\
		Red          & $2 083$       & $33.2 \%$ \\
		B$_G$        & $4 304$       & $2.6 \%$  \\
		B$_{pm}$     & $1 861$       & $0.9 \%$  \\
		R$_{pm}$     & $1 494$       & $2.7 \%$  \\
		\hline
	\end{tabular}
\end{table}

\section{Method}\label{sect:method}
\subsection{Disk model}\label{subsect:disk_model}
Given their brightness and their position in the CMD, the sources in the `fiducial' sample that have been selected are in principle bright blue young stars. As a result, they reside not far from their birthplace, which is most likely in the plane of the disk.
However, we know that stars in galaxies are not in a steady state: even young stars undergo heating, which causes them to deviate from circular orbits and causes asymmetric drift and additional disturbances, like those from spiral arms and vertical wobbling, that cause them to deviate from a circular trajectory. This results in a change in their velocity in terms of direction and norm. Unfortunately, we have no clue as to the actual position of a star in relation to the disk. We are then forced to model the velocities ($v_{\rm circ}$) as those belonging to a disk in perfect circular rotation with respect to the distance to the galaxy centre (radius R) considered ($v_{\rm circ} = f(R)$). But one of the advantages of using a survey like \egdrthree which covers the whole galaxy is that it is reasonable to assume that the majority of the perturbations statistically balance each other out.

\begin{figure}
	\includegraphics[width=\columnwidth]{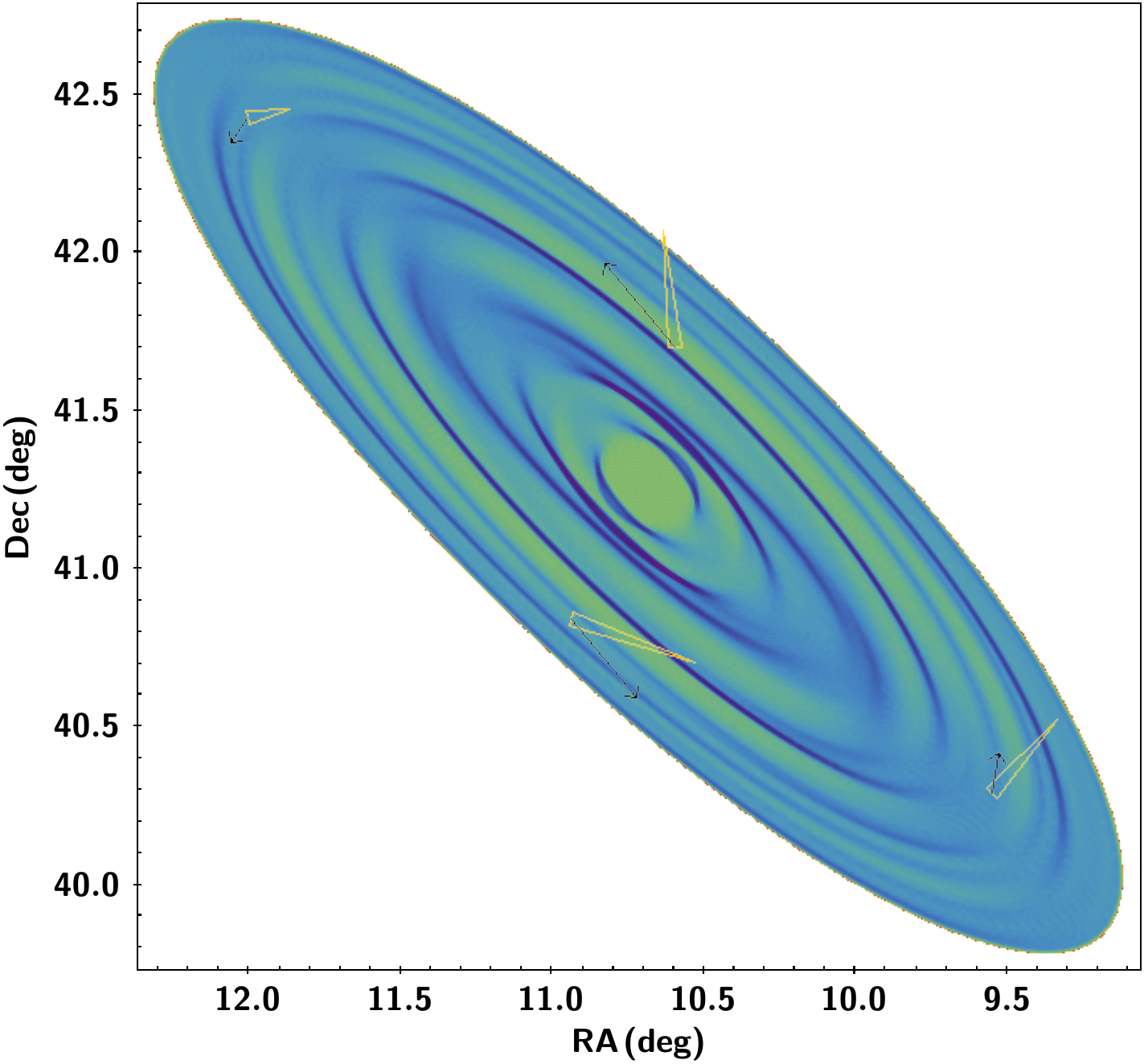}
    \caption{Disk model placed in the observed configuration of M31. Colour highlights the variation in projected density, which has been added purely as a visual aid to show the variation of $i$ and $\rm PA$ in this tilted ring model. Black arrows represent projected rotation of stars in the disk and yellow triangles, same velocities where the reflex displacement of the Sun is added. The scale between the two is preserved.}
    \label{fig:model_tourne_iPAvar_sun}
\end{figure}

From our position, the disk is observed with a tilt angle $i$ (angle between the line of sight and the plane) that varies with radius. This inclination takes place along an axis whose direction vector is oriented at the angle of position PA (angle between the north direction and the major axis of the projected ellipse) which also varies with radius. To model variations with respect to radius of these two angles, as well as those of the circular velocities, we use the values derived from $\rm{HI}$ observations by \cite{Chemin09}. They provide (in their Table 4) the best fit parameters with a tilted ring model to the HI rotation curve. We first model the disk face-on, up to a radius of $25 \kpc$, on a grid with a resolution of 1000 per axis. The model is then placed at the distance of M31 with the corresponding angles $i$ and $\rm PA$, and projected on the sky (see Figure~\ref{fig:model_tourne_iPAvar_sun}).

\subsection{Approach}

As we have seen in the previous section, stars are considered to move in circular motion on tilted rings. To this intrinsic rotation model, we add the bulk space motion of M31 that we aim to deduce, and project into proper motion observables $(\mu_{\alpha}, \mu_{\delta})$ in the heliocentric frame.

In order to fit our model with \egdrthree observations, we make use of a Markov Chain Monte Carlo (MCMC) method. The proper motion correlations between right ascension and declination given in the \egdrthree catalogue are also taken into account in the fitting process. The probability density function for a star to correspond to its modelled counterpart is given by:
\begin{align}\label{eq:proba}
\begin{split}
f(\mu_\alpha,\mu_\delta) & = {{1}\over{2\pi \sigma_{\mu_\alpha} \sigma_{\mu_\delta} \sqrt{1-\rho^2}}}  \times \\
& \exp \Bigg(-{{1}\over{2(1-\rho^2)}} \bigg[ {{\Delta^2_{\mu_\alpha}}\over{\sigma^2_{\mu_\alpha}}} + {{\Delta^2_{\mu_\delta}}\over{\sigma^2_{\mu_\delta}}}  
- {{2\rho \Delta_{\mu_\alpha} \Delta_{\mu_\delta}}\over{\sigma_{\mu_\alpha} \sigma_{\mu_\delta}   }}  \bigg] \Bigg)  \\
\end{split}
\end{align}
where $\Delta_{\mu_\alpha}$ and $\Delta_{\mu_\delta}$ are the measured offsets from the model. The Gaia proper motion uncertainties $\sigma_{\mu_\alpha}$ and $\sigma_{\mu_\delta}$ are thus taken into account along with their measured correlation $\rho \equiv {\tt pmra\_pmdec\_corr}$ (discussed in \citealt{Lindegren18}). 

As a reminder, this model effectively assumes a zero velocity dispersion of stars around their circular motion. This is obviously not correct, but the dispersion and asymmetric drift should be small for young stars and thus not bias our measurement based on the fiducial sample.

We model the effect of the contaminants by adding a term in the likelihood function:
\begin{equation}
    \ln{\cal L} = \sum_{k=1}^{n} \ln{\big[ (1-\eta_c)f(\mu_\alpha,\mu_\delta) + \eta_c g(\mu_\alpha,\mu_\delta)  \big]} \, ,
\end{equation}
where $\eta_c$ is the contamination fraction in the sample under consideration, and $g$ is the probability density function of the contaminants ($g$ is identical to Equation~\ref{eq:proba} but uses the proper motion mean and dispersion estimated for the contaminants of the particular sample). This likelihood formulation allows us to minimize the impact of sources not belonging to the M31 galaxy, and the large dispersions in the contaminating population decrease the influence of any bona fide M31 stars that have a deviant proper motion measurement.

\subsection{Validation}
The method is now tested to examine whether it is able to recover correct solutions given data with large uncertainties and contaminants. At the same time we test the influence of biases introduced by the spatial distribution of our sample which is not completely homogeneous, in terms of sky coverage and galactic radius.

To this end, fake observational data are built, based on our fiducial sample presented in Section~\ref{sect:selection}. We consider 1 919 points placed at the same position as those in our sample. We assigned to each of them the exact value of the rotation in the disk corrected for the solar reflex motion $(\mu_{\alpha}, \mu_{\delta})_{\rm disk}$, as defined in our model. We then add a random velocity to the model, drawn from a normal distribution centred on 0 with a dispersion of $150\muasyr$. This proper motion dispersion is deliberately very large in order to encompass all plausible situations of interest. The random velocity represents the centre of mass (COM) motion the method will have to recover. To model the proper motion errors, we add to each of the points a random proper motion on each direction $(\mu_{\alpha}, \mu_{\delta})_{\rm pec}$, randomly drawn in a normal distribution centred on 0 with a dispersion $(\sigma_{\alpha}, \sigma_{\delta})$ given by each data point. 

Finally, we choose a subset of the points to represent the contamination (1.8\% of the sample for the case of the fiducial sample). These points are reassigned proper motion values by drawing randomly from the function $g$ above.

\begin{figure}
	\includegraphics[width=\columnwidth]{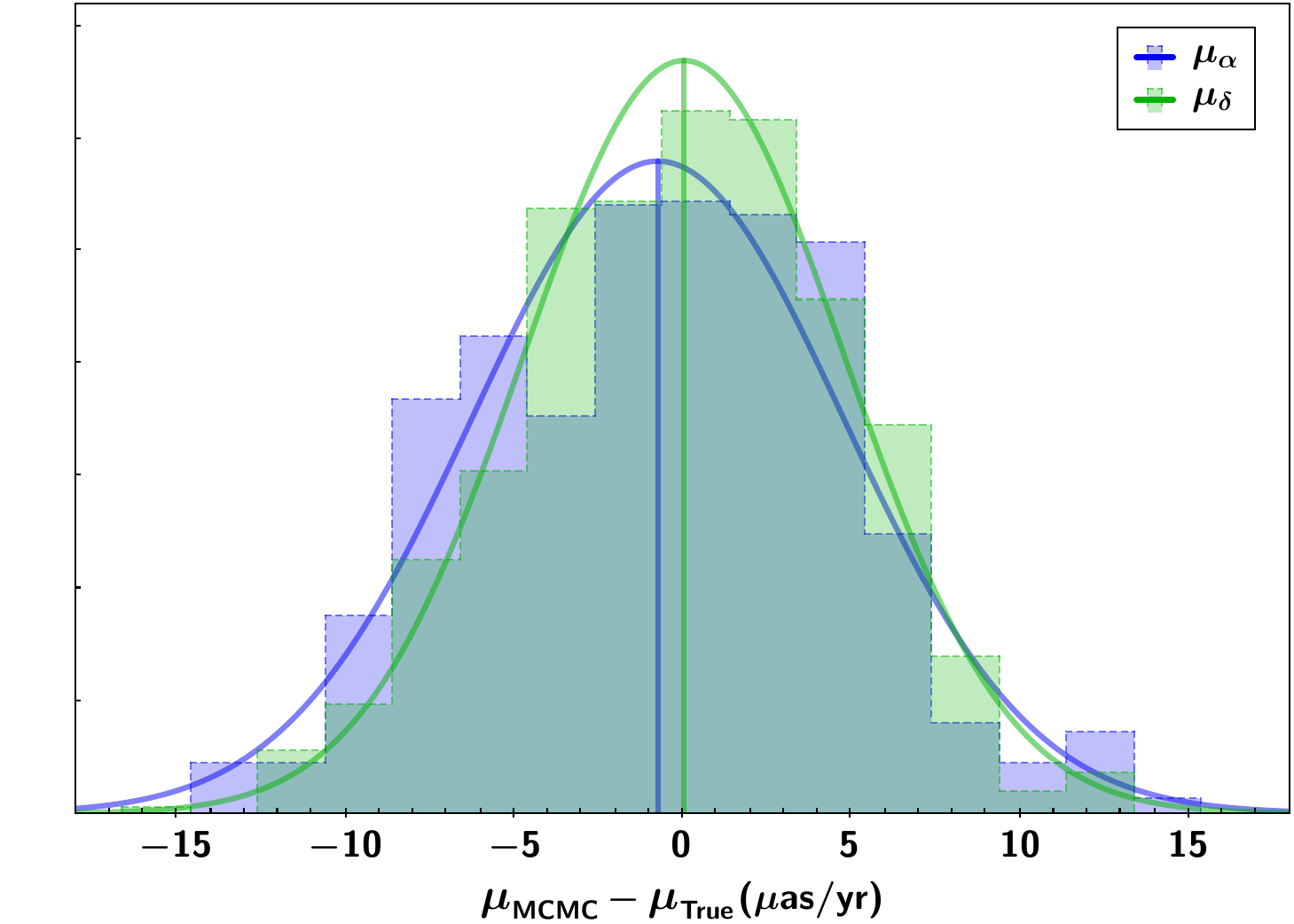}
    \caption{Normalised histogram of the differences in $\muasyr$ between results from the best-fitting method and the input value of the M31 COM given to the model to build fake data. The blue historgram shows the results for the right ascension direction and green shows the declination direction results. Simple Gaussian functions are superimposed to highlight the derived dispersions of the errors.}
    \label{fig:valid}
\end{figure}

We build a thousand such models of M31 and the contamination in the field, and each time we apply our fitting method in order to recover the proper motion of the COM. The difference between the value obtained and the value used to build the model indicates the systematic biases with their deviations induced both by the observed sample and by our method. This results in the values of $(\mu_{\alpha}, \mu_{\delta})_{\rm sys} = (-0.73 \pm 5.56, 0.36 \pm 4.49) \muasyr$ (see Figure~\ref{fig:valid}) for the fiducial sample. When applying the same test on the other samples, deviations are found to be of the same order of magnitude. These results show that our method does not produce strong biases in the measured bulk proper motion of M31.

\section{Results}\label{sect:results}
\subsection{Proper motion zero-point}

The \egdrthree catalogue goes far beyond the quality of the \gdrtwo catalogue, especially in terms of proper motion accuracy. However, there are still systematic measurement biases. For example, the offset in the right ascension direction is $10 \muasyr$ for bright sources ($G < 13$, \citet{Fabricius21}). But in this study, the sources considered are much fainter.

Thus we have decided to derive a local value of the proper motion zero points in each direction that we will take into account. To obtain them, we use quasars (QSOs), which due to their distance should have zero proper motion.
The \egdrthree sources being considered as quasars are selected within a radius of 20 degrees, centred around the position of M31. We used the QSOs listed in the table {\texttt{agn\_cross\_id}} published as part of the \textit{Gaia} archive. We also use the catalogue of quasars by \citet{Liao19}, which was especially compiled with the aim of being used for the validation of the {\it Gaia} mission. A cross-match between these two catalogues gives us 27,407 known QSOs.

\begin{figure}
	\includegraphics[width=\columnwidth]{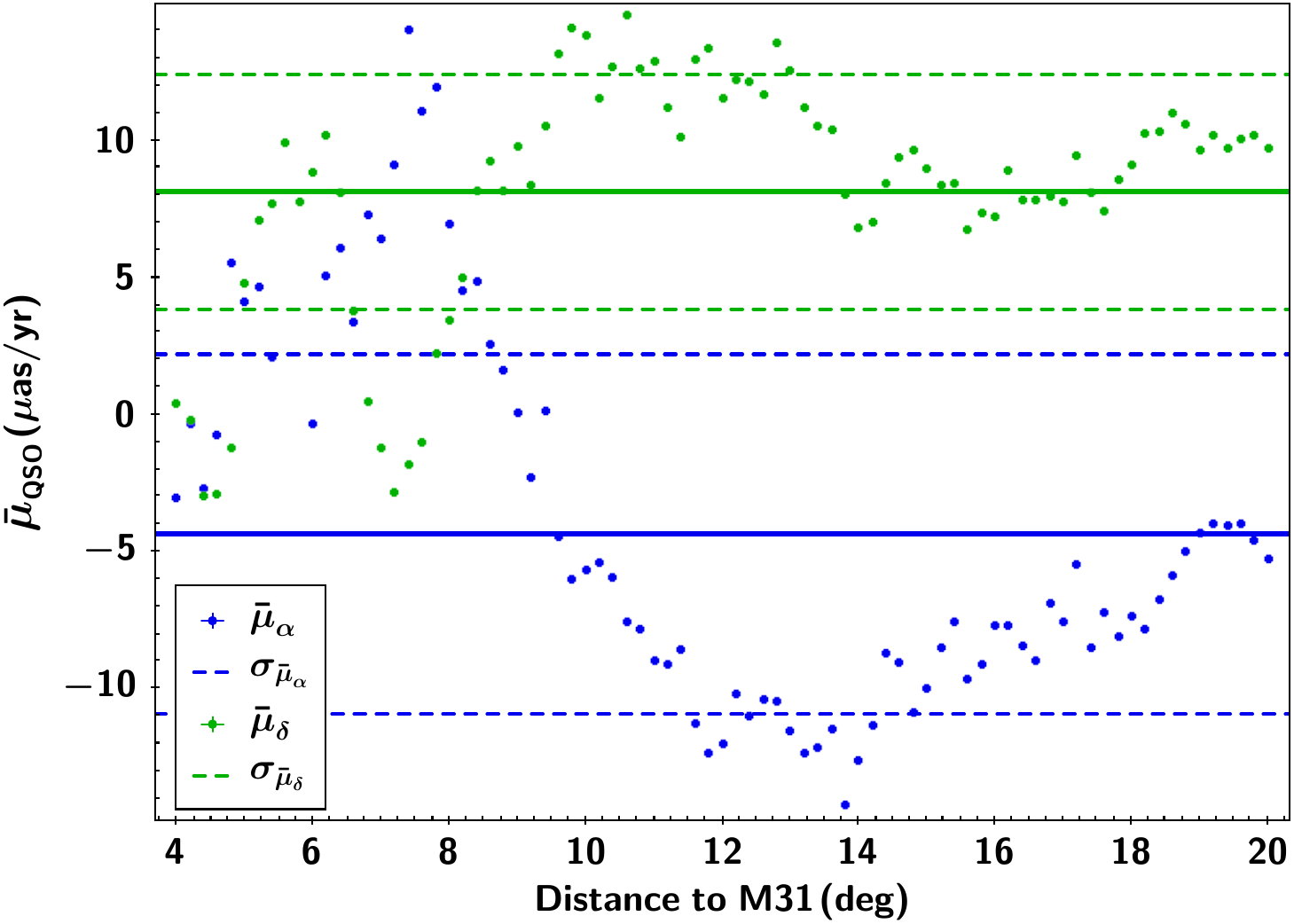}
    \caption{Average velocities of \egdrthree QSOs, common to the catalogue of \citet{Liao19}, in right ascension (blue) and declination (green) as a function of the angular distance to the centre of the Andromeda galaxy. The points represent the average apparent motions of the objects located between the centre and the given radius. The solid lines are the respective averages of all the circular proper motion averages and the dashed lines their $1\sigma$ dispersions.}
    \label{fig:QSOs_pmMean}
\end{figure}

We derive the difference in proper motion between these two catalogues in concentric circles centred on M31 COM, using radii from 4$\deg$ up to 20$\deg$ (see Figure~\ref{fig:QSOs_pmMean}).
The lower limit is set by the radius at which a circle centred on M31 contains at least 1000 QSOs, which seems to us to be a sufficient value to have a reliable statistical value. The overall mean as well as the average dispersion are then calculated to get the errors and uncertainties caused by the zero point offset of the proper motions, $(\mu_{\alpha}, \mu_{\delta})_{\rm off} = (-4.36 \pm 6.56, 8.13 \pm 4.28) \muasyr$.

\subsection{Andromeda proper motion}

The method developed in this study is applied to each of the selections described above to obtain the proper motions (and the corresponding transverse velocities assuming a distance of $785\kpc$) in the heliocentric frame. These raw values are listed in Table~\ref{tab:trans_vel}. Correcting these values for the proper motion zero-points and for the small systemic bias (deduced from the method validation tests) yields the values given in Table~\ref{tab:trans_velQSO}. 

Lastly, the proper motions and velocities are converted into a non-rotating reference frame centered at the position of the Sun.
To this end we make use of the recent value of the motion of the Sun with respect to the Galactic centre derived by \cite{Reid19}, $(U_\odot, V_\odot+V_c, W_\odot) = (10.6 \pm 1.2, 247 \pm 4, 7.6 \pm 0.7) \kms$. These yield a reflex proper motion of $(\mu_{\alpha}, \mu_{\delta})_{\odot} = (37.6 \pm 0.6, -20.9 \pm 0.4) \muasyr$. When subtracting $\mu_{\odot}$ from the corrected values, we thus derive the final transverse velocity of M31 with respect to the centre of the Milky Way. These values are provided in Table~\ref{tab:trans_velMW}.
To evaluate the impact of the choice of the solar motion on the final values, several other solar velocity values given in the literature were also considered but the differences on the transverse velocity of M31 are always smaller than $5 \kms$ in each direction. Hence, the velocity of the Sun has a moderate impact on our results, given the current uncertainties on the determination of the proper motion of M31.

\begin{figure}
	\includegraphics[width=\columnwidth]{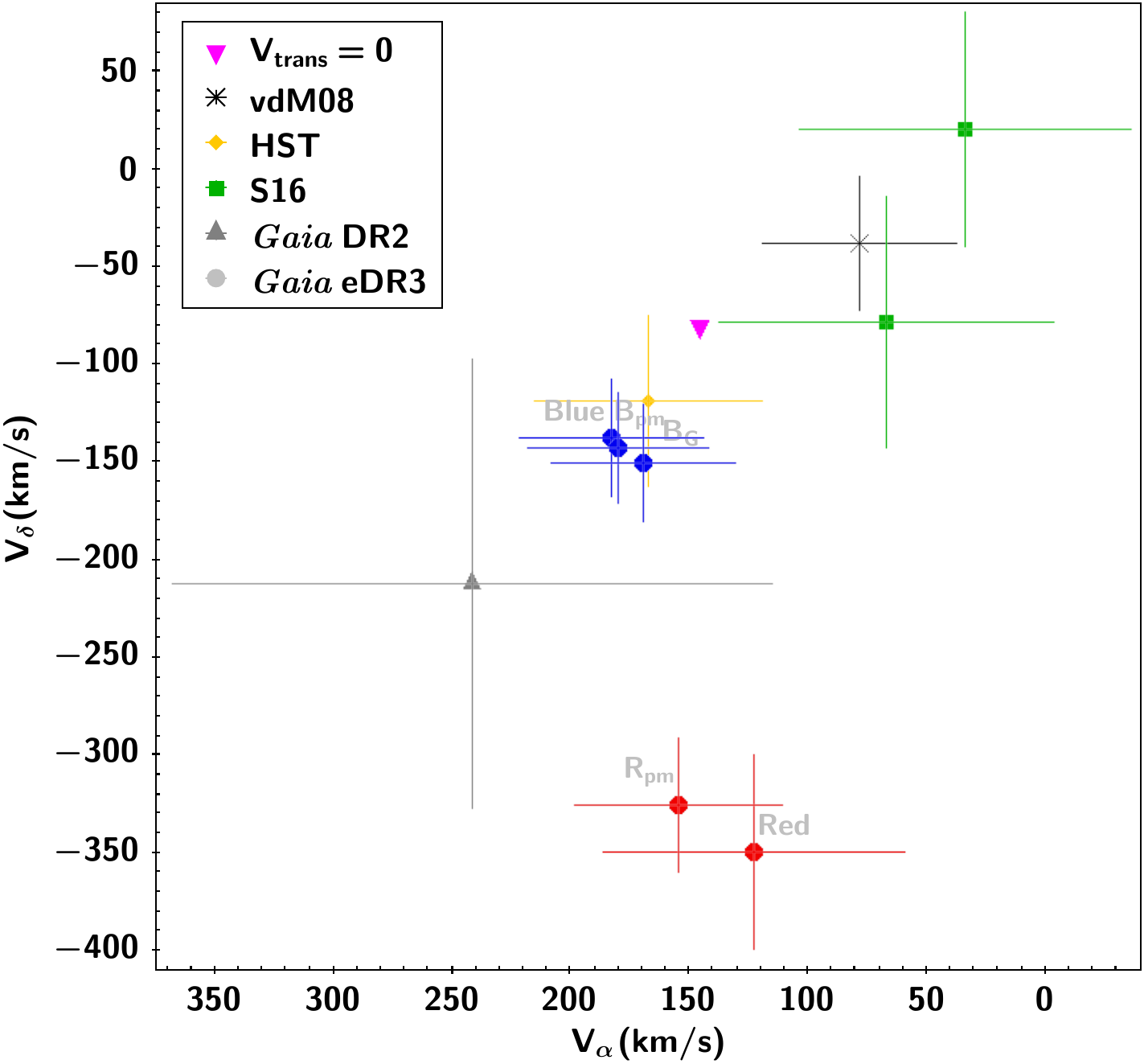}
    \caption{Transverse velocities in the heliocentric frame with their associated uncertainties. The purple triangle indicates a strictly radial approach between M31 and the MW. The black cross is the weighted proper motion of 4 results based on LOS velocity of satellite galaxies by \citet{vdM08}. The yellow diamond is the weighted proper motion of three fields observed with HST \citep{Sohn12, vdM12a}. The green triangles are estimates deduced from the perspective motion of satellite galaxies (and hence derived from radial velocity measurements), including (upper right) or not the plane of satellites \citep{Salomon16}. The grey triangle is the recent value derived from the \gdrtwo catalogue \citep{vdM19}. Blue and red points are results derived in this study with \egdrthree data for blue young main sequence stars and red super-giant stars (respectively) using the different samples detailed in Section~\ref{sect:subsample}.}
    \label{fig:velSummary}
\end{figure}

\begin{table*}
	\centering
	\caption{Transverse velocities in the \egdrthree frame.}
	\label{tab:trans_vel}
	\begin{tabular}{lcccccc} 
		\hline
		Sample & $\mu_\alpha$ & $\mu_\delta$ & $\mu$ & $V_{\alpha}$ & $V_{\delta}$ & $V_{\rm Trans}$\\
		       & $\muasyr$ & $\muasyr$ & $\muasyr$ &  $\kms$      & $\kms$       & $\kms$       \\
		\hline
		
		Blue & $    43.8 \pm    6.1 $ & $   -28.4 \pm    5.2 $ & $    52.5 \pm    5.8$ & $   163.1 \pm   22.7 $ & $  -105.6 \pm   19.5 $ & $   195.4 \pm   21.7$ \\

		B$_{\rm G}$ & $    40.4 \pm    5.9 $ & $   -32.0 \pm    5.2 $ & $    51.8 \pm    5.6$ & $   150.5 \pm   22.0 $ & $  -119.0 \pm   19.2 $ & $   192.9 \pm   20.9$ \\

		B$_{\rm pm}$ & $    42.9 \pm    5.6 $ & $   -29.9 \pm    4.7 $ & $    52.5 \pm    5.3$ & $   159.8 \pm   20.7 $ & $  -111.2 \pm   17.4 $ & $   195.5 \pm   19.7$ \\

		\\
        Red & $    27.7 \pm   14.8 $ & $   -85.1 \pm   11.9 $ & $    90.7 \pm   12.2$ & $   103.1 \pm   55.0 $ & $  -316.8 \pm   44.4 $ & $   337.5 \pm   45.4$ \\

        R$_{\rm pm}$ & $    36.2 \pm    8.0 $ & $   -78.9 \pm    6.8 $ & $    87.2 \pm    7.0$ & $   134.8 \pm   29.9 $ & $  -293.6 \pm   25.2 $ & $   324.4 \pm   26.1$ \\

		\hline
	\end{tabular}
\end{table*}

\begin{table*}
	\centering
	\caption{Transverse velocities corrected for apparent QSO motion and systematic method bias.}
	\label{tab:trans_velQSO}
	\begin{tabular}{lcccccc} 
		\hline
		Sample & $\mu_\alpha$ & $\mu_\delta$ & $\mu$ & $V_{\alpha}$ & $V_{\delta}$ & $V_{\rm Trans}$\\
		       & $\muasyr$ & $\muasyr$ & $\muasyr$ &  $\kms$      & $\kms$       & $\kms$       \\
		\hline
		
		Blue & $    48.9 \pm   10.5 $ & $   -36.9 \pm    8.1 $ & $    61.9 \pm    9.7$ & $   182.1 \pm   39.2 $ & $  -137.2 \pm   30.2 $ & $   230.5 \pm   35.9$ \\

		B$_{\rm G}$ & $    45.5 \pm   10.4 $ & $   -40.5 \pm    8.1 $ & $    61.6 \pm    9.4$ & $   169.4 \pm   38.8 $ & $  -150.5 \pm   30.0 $ & $   229.2 \pm   34.9$ \\

		B$_{\rm pm}$ & $    48.0 \pm   10.2 $ & $   -38.4 \pm    7.8 $ & $    62.1 \pm    9.3$ & $   178.7 \pm   38.1 $ & $  -142.7 \pm   28.9 $ & $   231.1 \pm   34.6$ \\

		\\
        Red & $    32.8 \pm   17.1 $ & $   -93.6 \pm   13.5 $ & $   100.6 \pm   13.8$ & $   122.0 \pm   63.7 $ & $  -348.4 \pm   50.1 $ & $   374.4 \pm   51.5$ \\

        R$_{\rm pm}$ & $    41.3 \pm   11.8 $ & $   -87.4 \pm    9.2 $ & $    97.3 \pm    9.7$ & $   153.7 \pm   43.8 $ & $  -325.1 \pm   34.2 $ & $   362.1 \pm   36.0$ \\

		\hline
	\end{tabular}
\end{table*}

\begin{table*}
	\centering
	\caption{Transverse velocities in non-rotating Galactic frame, centered on the Sun.}
	\label{tab:trans_velMW}
	\begin{tabular}{lcccccc} 
		\hline
		Sample & $\mu_\alpha$ & $\mu_\delta$ & $\mu$ & $V_{\alpha}$ & $V_{\delta}$ & $V_{\rm Trans}$\\
		       & $\muasyr$ & $\muasyr$ & $\muasyr$ &  $\kms$      & $\kms$       & $\kms$       \\
		\hline

		Blue & $    11.3 \pm   10.6 $ & $   -16.0 \pm    8.1 $ & $    22.1 \pm    8.4$ & $    42.2 \pm   39.3 $ & $   -59.4 \pm   30.3 $ & $    82.4 \pm   31.2$ \\

		B$_{\rm G}$ & $     7.9 \pm   10.5 $ & $   -19.6 \pm    8.1 $ & $    23.6 \pm    8.1$ & $    29.5 \pm   38.9 $ & $   -72.8 \pm   30.0 $ & $    87.6 \pm   30.1$ \\

		B$_{\rm pm}$ & $    10.4 \pm   10.3 $ & $   -17.5 \pm    7.8 $ & $    22.7 \pm    8.0$ & $    38.8 \pm   38.2 $ & $   -65.0 \pm   29.0 $ & $    84.5 \pm   29.9$ \\

		\\
        Red & $    -4.8 \pm   17.1 $ & $   -72.7 \pm   13.5 $ & $    74.9 \pm   13.4$ & $   -17.9 \pm   63.7 $ & $  -270.6 \pm   50.1 $ & $   278.6 \pm   49.9$ \\

        R$_{\rm pm}$ & $     3.7 \pm   11.8 $ & $   -66.5 \pm    9.2 $ & $    67.6 \pm    9.2$ & $    13.8 \pm   43.9 $ & $  -247.4 \pm   34.2 $ & $   251.6 \pm   34.1$ \\

		\hline
	\end{tabular}
\end{table*}

\section{Red sample contamination}\label{sect:red_Discussion}

Despite the smaller number of estimated bona fide red stars ($\sim$1391) compared to blue stars ($\sim$1884), our MCMC method is still able to measure the proper motions with small uncertainties, as can be seen in Tables~1--3. However, the derived motions of the red samples (red points in Figure~\ref{fig:velSummary}) are considerably different to those of the blue samples. We ran several tests to attempt to measure this offset in a differential manner, exploring the central disk, the inner and outer rings and portions thereof. Furthermore, we also tried different disk models and various contamination levels. In all of the tests we conducted, including when using spatially-selected sub-samples throughout M31, this proper motion offset was reproducible within the derived uncertainties. We also examined M31's satellite galaxy M33 using identical colour cuts and found good agreement between the blue and red samples there.
The discrepancy therefore does not appear to be intrinsic to {\it Gaia}, as might have occurred if there had been a strong colour-dependent proper motion bias.
We suspect instead that the offset is due to a combination of factors including foreground contamination.

\subsection{Galactic model}
In order to explore this option more in detail, we make use of the {\it Gaia} Universe Model snapshot (GUMS, \citealt{Robin12}) based on the Besançon Galactic Model. It represents the state of the art of our present knowledge about the Milky Way. It is very useful in our case as it aims to statistically reproduce, among other things, the star counts and velocities of Galactic sources along the line of sight. Simulations of the expected {\it Gaia} catalogues based on GUMS are produced using the {\it Gaia} Object Generator (GOG, \citealt{Luri14}). Hence, we have at our disposal an accurate mock {\it Gaia} catalogue of the Milky Way, which does not include streams, dwarf galaxies, globular clusters and M31. The same observable data as in the \egdrthree catalogue are available, in particular colours, magnitudes, proper motions as well as the input parameters used in the simulation like spectral type of stars or the dynamical population which the star belongs to.

\subsection{Foreground selection}
We then apply to this GOG catalogue the same procedure as in section~\ref{sect:selection} to retain stars around the M31 line of sight within identical spatial selections and in the same CMD selections. With the blue cuts, 19 stars are contained in GOG within the central ellipse and 811 with the red cuts. That corresponds to a contamination of about 1\% and 38.9\% respectively when comparing to the observed number of blue and red sources in this area. The percentages are qualitatively of the same order as the estimated contamination we evaluated with the \egdrthree control samples thanks to the north-west and south-east ellipses (see column 2 in Table~\ref{tab:sample_sum}). Sources in the GOG catalogue are mainly late K and M dwarf stars. Within the field of 2 degrees centered at the  position of Andromeda, the GOG blue sample is composed of 24\% of stars from the thin disk, 65\% from the thick disk and 11\% from the halo whereas the GOG red sample contains 63\%, 36\% and 1\% in these components respectively.

\begin{figure}
	\includegraphics[width=\columnwidth]{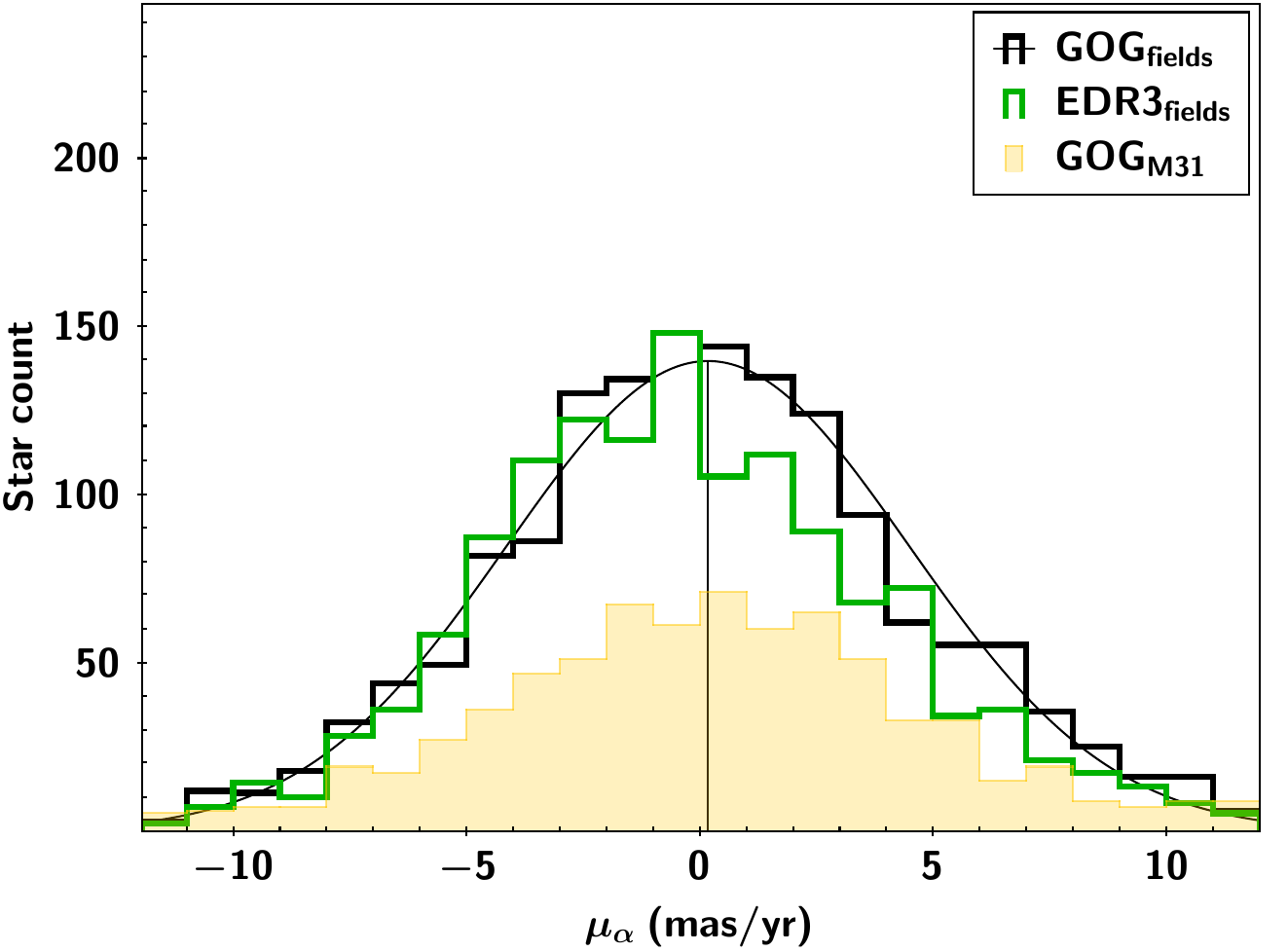}
	\includegraphics[width=\columnwidth]{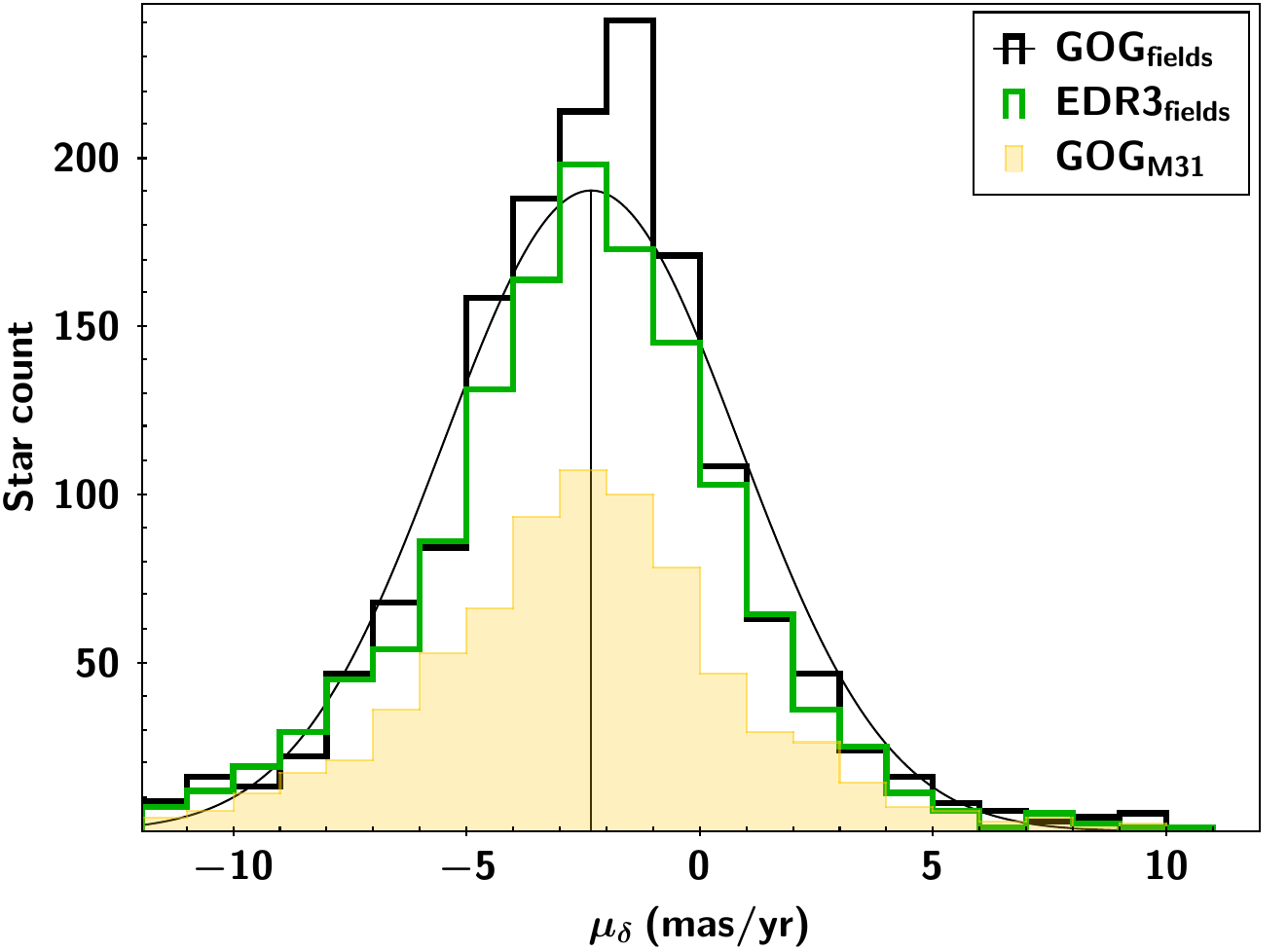}
    \caption{Distribution of proper motion in
    \masyr{} along the right ascension (top panel) and in the declination direction (bottom panel) for samples of red sources. The black and green histograms are the distribution of sources within the two joined elliptical north-west and south-east control fields, for the GOG catalogue and \egdrthree respectively. The yellow distribution is that of the central ellipse for the GOG catalogue. Note that the amount of sources for the black and green distributions (two ellipses each) has to be divided by two when comparing with the yellow one which covered only half of the field in the sky (one ellipse). }
    \label{fig:GOG_pm}
\end{figure}

These stars follow the velocity dispersion of Galactic stars and they are close to the observer. Consequently the distribution of their proper motions is scattered over a large range, compared to the distribution of M31 sources. Moreover, as most of the stars belong to the disks, combined with the fact we are observing in a specific direction, they have correlated kinematics due to the rotation of the Galaxy. The proper motion distribution of the two joined (north-west plus south-east ellipses) GOG red samples is $(\mu_{\alpha}, \mu_{\delta})_{\rm{GOG} \, fields} = (0.15 \pm 4.32, -2.35 \pm 3.17) \masyr$. This has to be compared with the two joined \egdrthree red samples in the same areas $(\mu_{\alpha}, \mu_{\delta})_{\rm{EDR3 \, fields}} = (0.22 \pm 5.31, -2.66 \pm 3.42) \masyr$ (see Figure~\ref{fig:GOG_pm}). 

Several conclusions can be drawn from this comparison with the GOG simulation. First, we have identified the origin of the main source of contaminants which is indeed Milky Way disk dwarf stars. The level of contamination is broadly consistent with our estimation from the \egdrthree control samples both for blue and red samples. Given the very small number of blue sources, we cannot extrapolate more on this sample other than confirming that the contamination is very low. It seems however that there are slightly more red contaminating stars in the GOG catalogue. This could be due to the fact that there is no crowding cut-off in the GOG star selection. Second, the GOG and \egdrthree red samples in the control fields have similar values of the proper motion average and dispersion (in both proper motion directions) which confirms that the kinematic behaviour of the stars in this field is correctly predicted by the GOG catalogue (see Figure~\ref{fig:GOG_pm}). Third, the global offset in proper motion of the contaminant distribution in the red selection follows the same trend as is observed for the red population of M31: namely almost no deviation of the central value in the right ascension direction but a large shift towards the south. Of course, the proper motion of the contaminant population and the derived proper motion for M31 are not directly comparable as there is an amplitude difference of two order of magnitudes. Nevertheless, the trends are similar and strongly suggest that the presence of Milky Way contaminant stars could bias results based on red samples.  


\subsection{Discussion}
Despite the fact that we now better understand the contamination of the red sample by the Milky Way foreground, there are still unresolved questions. We have attempted many stricter CMD selections on the red sample, but find that the discrepancy still remains between the proper motion of M31 derived with the blue samples and the red samples.
Several hypotheses can be envisaged without being confirmed with the current data.

Probably the simplest explanation is that the residual contamination affects the proper motion measurement of the red population most likely causing a bias towards the southern direction (Figure~\ref{fig:GOG_pm}).

The extinction could be much stronger than expected if some dust is located just on the M31 line of sight. This would change the CMD and bias our selection for the red samples, taking away bona fide stars and adding foreground stars. But when looking at dust content and structures with Planck data in the region around M31, there is no hint of such a configuration even if we cannot rule out this possibility.  

There may also exist a second source of contamination that cannot be identified with the current data. This could be for example from a stream.
The region around M31 contains several Galactic stellar streams, such as the ``PAndAS-MW'' stream identified by \citet{Martin14} (see their Figures 2 and 3), which passes in a $>1\deg$-wide band in front of M31. This would give rise to varying levels of foreground contamination, that would not be present in the blue sample, because old streams do not contain stars bluer than the main sequence turnoff (ignoring faint white dwarfs). Since we measure a large declination-direction proper motion offset of $(\Delta \mu_{\alpha}, \Delta \mu_{\delta}) = (6.7, 49.0) \muasyr$ between the blue and the red samples in the two central regions (using the proper motion filter), this could be a possible explanation even if we cannot confirm it.

\section{Conclusion}\label{sect:ccl}

In this study, we constructed five samples from \egdrthree with different selection criteria so as to best examine the evidence for the proper motion of M31. The fiducial sample (Blue) contains young main sequence stars with $G_{\rm BP} - G_{\rm RP} < 0.5$ and has a faint magnitude cut (retaining only stars with $16 < G < 20$). A more extended blue sample (B$_G$) includes the fainter stars. The red sample (Red) contains red super-giants. We also created two additional selections using a 3-sigma cut on proper motion, yielding a further blue (B$_{\rm pm}$) and red (R$_{\rm pm}$) sample.

The blue main sequence stars are brighter and more numerous, and the derived transverse velocities calculated from the blue samples cluster around the previously-measured weighted average HST value, as we show in Figure~\ref{fig:velSummary} (blue points).

With the advent of {\it Gaia} EDR3, we are now able to measure very accurate proper motions for some of the most distant bodies in the Local Group. In M31 we can now have at our disposal a panoramic view of the stellar motions over the whole body of the galactic disk, superseding earlier efforts with HST in small fields. Our final estimate results in a relative velocity vector between MW and M31 which is very close to being radial, in accordance with the \citet{vdM12a} weighted data with HST. It is therefore almost certain that the Local Group is gravitationally bound.

\section*{Acknowledgements}
The authors thank the anonymous referee for a thoughtful report. JBS would like to thank Julien Montillaud and Annie Robin for insightful conversations. RI and BF acknowledge funding from the Agence Nationale de la Recherche (ANR project ANR-18-CE31-0006, ANR-18-CE31-0017 and ANR-19-CE31-0017), from CNRS/INSU through the Programme National Galaxies et Cosmologie, and from the European Research Council (ERC) under the European Unions Horizon 2020 research and innovation programme (grant agreement No. 834148). NIL acknowledges financial support of the Project IDEXLYON at the University of Lyon under the Investments for the Future Program (ANR-16-IDEX- 0005). 

\section*{Data Availability}

This work has made use of publicly available data from the European Space Agency (ESA) mission {\it Gaia} (\url{https://www.cosmos.esa.int/gaia}), processed by the {\it Gaia} Data Processing and Analysis Consortium (DPAC,
\url{https://www.cosmos.esa.int/web/gaia/dpac/consortium}). Derived data products will be shared on reasonable request to the corresponding author.



\bibliographystyle{mnras}
\bibliography{biblio_M31PM} 





\bsp	
\label{lastpage}
\end{document}